\documentclass[twocolumn,preprintnumbers,
nofootinbib,prd]{revtex4}
\usepackage{graphicx}

\usepackage[hypertex]{hyperref}
\newcommand{\beq}{\begin{equation}}
\newcommand{\eeq}{\end{equation}}

\begin{document}

% page numbers bottom-center
\pagestyle{plain}

\preprint{MADPH-07-1500\,$\;$\\
          IZTECH-P-07/05\,$\;$\\
          DESY-07-213}

\title{Dirac Neutrino Masses from Generalized Supersymmetry Breaking}

%\author{Author list goes here..}
\author{Durmu{\c s}~A.~Demir\footnote{\tt email: durmus.demir@desy.de}}
\affiliation{Department of Physics, Izmir Institute of Technology, IZTECH,
TR35430 Izmir, Turkey\\
Deutsches Elektronen - Synchrotron, DESY, D-22603 Hamburg, Germany}

\author{Lisa~L.~Everett\footnote{\tt email: leverett@wisc.edu}}
\affiliation{Department of Physics, University of Wisconsin,
Madison, WI 53706, USA}

\author{Paul~Langacker\footnote{\tt email: pgl@ias.edu}}
\affiliation{School of Natural Sciences, Institute for Advanced Study, Einstein Drive,
Princeton, NJ , 08540, USA}

%%%%%%%%%%%%%%%%%%%%%%%%%%%%%%%%%%%%%%%%%%%%%%%%%%%%%%%%%%%%%%%%%%%%%%%%%%%%

\begin{abstract}
We demonstrate that  Dirac neutrino masses in the experimentally preferred range are generated within supersymmetric gauge extensions of the Standard Model with a generalized supersymmetry breaking sector.  If the usual superpotential Yukawa couplings are forbidden by the additional gauge symmetry (such as a $U(1)^\prime$), effective  Dirac mass terms involving the ``wrong Higgs" field can arise either at tree level due to hard supersymmetry breaking fermion Yukawa couplings, or at one-loop due to nonanalytic or ``nonholomorphic"  soft supersymmetry breaking trilinear scalar couplings.  As both of these operators are naturally suppressed in generic models of supersymmetry breaking, the resulting neutrino masses are naturally in the sub-eV range.   The neutrino magnetic and electric dipole moments resulting from the radiative mechanism also vanish at one-loop order. 
\end{abstract}
\maketitle

%%%%%%%%%%%%%%%%%%%%%%%%%%%%%%%%%%%%%%%%%%%%%%%%%%%%%%%%%%%%%%%%%%%%%%%%%%%
The discovery of neutrino oscillations has confirmed that
neutrinos are massive and that leptons exhibit nontrivial mixing,
providing the first particle physics evidence for physics beyond
the Standard Model (SM).  Neutrino masses
require either the existence of novel matter species not found in
the SM spectrum 
%at low energies 
and/or the violation of the global
symmetries of the SM via higher-dimensional operators.  Extensions incorporating such additional structure should ideally be capable of improving the ultraviolet behavior of the SM
beyond Fermi energies.  Low-energy softly-broken
supersymmetry thus provides a well-motivated theoretical framework in which to incorporate neutrino mass generation mechanisms.
%which to investigate neutrino mass generation and other aspects of
%physics beyond the SM.
%, as low-energy supersymmetry stabilizes the
%SM Higgs sector against quadratic divergences and can provide a cold dark matter candidate in
%the lightest neutralino if R-parity is conserved. 
 %Supersymmetry, however, does not imply
%anything special for massive neutrinos, and hence all that one
%can do is to incorporate neutrino mass generation
%mechanisms within a supersymmetric framework.  
As no conclusive
experimental indications for neutrinoless double beta decay (for Majorana neutrinos) or neutrino magnetic or electric
dipole moments (for Dirac neutrinos) are available at present, we must explore all mechanisms for generating light neutrino masses, not only to reveal the origin of neutrino masses and
mixings, but also to determine viable patterns for physics beyond
the SM.

Many mechanisms are known for generating light Majorana or Dirac neutrino masses (see e.g.~\cite{seesaw,weinberg, alter,models}). Some scenarios, including the familiar seesaw mechanism \cite{seesaw}, rely upon the supposition that the right-handed neutrinos have no gauge quantum numbers with respect to the low energy gauge group.    
%For example, in the seesaw mechanism this feature
%allows for potentially large right-handed neutrino mass terms
%which are unrelated to the spontaneous breaking of the electroweak
%or other symmetries.   
Right-handed neutrinos are SM gauge singlets, but they can be charged under additional gauge symmetries which may survive from many high-scale theories, such as four-dimensional string models.
Thus, if the right-handed neutrinos are not complete (low scale) gauge singlets, these scenarios are not viable, at least not in their simplest implementation.  %A notable exception is the case
Dirac neutrinos, which occur if lepton number is an exact symmetry, do not necessarily have this requirement.  However, as neutrino Dirac masses originate from Yukawa interactions after electroweak breaking, their Yukawa couplings must be exceedingly small. This can be explained if they are forbidden at the renormalizable level by additional symmetries but are generated from higher-dimensional operators. Previous work along these lines (see e.g.~\cite{higherd}) assumes that such operators occur in the superpotential.

In this Letter, we demonstrate that appropriately suppressed Dirac neutrino masses can be generated 
by generalized supersymmetry breaking
terms in models in which the right-handed neutrinos are charged under additional gauge symmetries.  These symmetries forbid the usual neutrino superpotential Yukawa couplings, but allow for higher-dimensional operators which lead to effective Dirac neutrino mass terms involving the ``wrong Higgs" field upon supersymmetry breaking.
%The ingredients and implications of the mechanism are as follows.

Fermion masses represent the breakdown of chiral flavor
symmetries, and as such can be parametrized by vacuum expectation
values (VEVs) of scalar fields charged under the flavor symmetry.
In theories with low energy  supersymmetry, it has
long been known \cite{soft1} (see also \cite{kane}) that such chiral flavor
symmetries may be broken by the VEVs of auxiliary fields, rather than their
scalar counterparts.  If the renormalizable superpotential Yukawa couplings and right-handed neutrino Majorana mass terms are forbidden by the symmetry, fermion masses are generated either
 (i) at tree level due to hard supersymmetry breaking effective Yukawa terms, with
\begin{equation}
\label{treemass1}
m_f\sim Y_{{\rm eff}} \langle H \rangle ,
%\sim \frac{\tilde{m}}{M} \langle H \rangle,
\end{equation}
 or (ii) radiatively via sfermion--neutralino loops: % as shown schematically in Figure 1. 
%\cite{soft1}
\begin{equation}
\label{radmass}
m_f\sim \frac{\alpha}{2\pi}\frac{\tilde{A}M_\lambda \langle H \rangle}{\tilde{m}^2},
\end{equation}
in which $\alpha$ denotes a typical gauge coupling, $\tilde{A}$ denotes a soft trilinear scalar coupling, $M_\lambda$ denotes a gaugino mass, and 
$\tilde{m}$ denotes a typical sfermion mass.

In generic supersymmetry breaking models, Eq.~(\ref{treemass1}) and Eq.~(\ref{radmass}) are naturally in the experimentally favored ranges for 
neutrino masses.  The effective Yukawa interaction of Eq.~(\ref{treemass1}) is due to a higher-dimensional K\"{a}hler potential operator suppressed by a 
high scale $M$ (the messenger scale).  Hence, $Y_{{\rm eff}}\sim \tilde{m}/M$, and
\begin{equation}
\label{estimate1}
\left (\frac{m_\nu}{10^{-3}\,{\rm eV}}\right )\sim \left (\frac{\tilde{m}}{100 \,{\rm GeV}}\right )\left (\frac{M}{10^{16}\,{\rm GeV}} \right )^{-1}.
\end{equation}
Due to the large suppression factor, these effective ``wrong-Higgs" Yukawa coupling terms do not spoil the resolution to the hierarchy problem, although technically they are hard supersymmetry breaking operators \cite{seiberg}.

Let us now focus on the radiative mass terms of Eq.~(\ref{radmass}).  These terms are suppressed due to the specific trilinear scalar couplings (the $\tilde{A}$ terms) allowed by the flavor symmetry.  To understand this suppression,  recall that there are two classes of $\tilde{A}$ terms:  (i) the standard analytic or
``holomorphic" terms, which are coefficients of operators of the form 
%\begin{equation}
%\label{hol}
%-\mathcal{L}=A
$\phi\phi \phi$, 
%\end{equation} 
and (ii) the 
nonanalytic or ``nonholomorphic" terms, which accompany  $\phi^*\phi\phi$ operators.
%\begin{equation}
%\label{nonhol}
%-\mathcal{L}=A^\prime 

%\end{equation}

In typical models, the nonanalytic trilinear scalar terms, which have previously been considered in the context of radiative SM fermion mass generation \cite{soft1}, are well known to be suppressed by $\tilde{m}/M$ \cite{nonhol,seiberg}.  Recently, it has been claimed that without this strong suppression, Goldstino loops can reintroduce the hierarchy problem \cite{bagger}.  If these terms are so strongly suppressed, they are irrelevant for most phenomenological analyses and cannot provide the dominant contribution to charged fermion masses.  However, this suppression is of the right order to be relevant for Dirac neutrino masses:  %Qualitatively, they are of the form
\begin{equation}
\label{estimate2}
\left (\frac{m_\nu}{10^{-3}\,{\rm eV}}\right )\sim \frac{\alpha}{2\pi}\left (\frac{\tilde{m}}{100 \,{\rm GeV}}\right )\left (\frac{M}{10^{16}\,{\rm GeV}} \right )^{-1},
\end{equation}
which can fall within the experimentally
allowed range without excessive tuning. Furthermore, the associated radiative neutrino 
magnetic and electric dipole moments vanish at one
loop level. 

The nonanalytic terms contribute to quadratic divergences through tadpole diagrams \cite{softterm}, and thus are not soft in the presence of gauge 
singlets.  If SM singlets such as right-handed neutrinos are present  these terms can be rendered soft only if the SM gauge group is extended, and 
all SM singlets are charged under the additional gauge group(s). The simplest extension is to include an additional Abelian $U(1)^{\prime}$ factor, 
which can also provide a resolution of the supersymmetric $\mu$ problem \cite{u1p}. The $U(1)^{\prime}$ charges can be assigned such that the neutrino 
superpotential Yukawa couplings and the associated trilinear K\"{a}hler potential couplings are forbidden, while the wrong-Higgs trilinear couplings 
are allowed.  The nontrivial $U(1)^\prime$ charges of the right-handed neutrinos also forbid bare Majorana mass terms.

We will now provide a detailed analysis of these points.  Consider the MSSM
augmented by three right-handed neutrino superfields, $\widehat{N}^i
= \left ( \widetilde{\nu}_R^{c\,i}, \nu_R^{c\, i} \right)$.  Supersymmetry breaking occurs in a hidden
sector via the $F$-component VEV
of a chiral superfield $\widehat{X}$, with $\langle
\widehat{X} \rangle = F \theta \theta$, and is
communicated to the visible sector at a large scale $M$ via
nonrenormalizable interactions.  The $F$ component of
the neutrino superpotential Yukawa coupling then gives an analytic scalar
trilinear coupling:\footnote{The extension to quarks and charged leptons is straightforward.}
\begin{eqnarray}
\label{anu} \frac{1}{M}\left(\widehat{X} \widehat{L} \cdot
\widehat{H}_u {\bf Y_{\nu}}  \widehat{N}\right)_{F}\, = \,
\widetilde{L} \cdot {H}_u {\bf A_{\nu}}
\widetilde{\nu}_R^{c},
\end{eqnarray}
with
\begin{eqnarray}
 {\bf A_{\nu}}
\equiv \frac{F}{M} {\bf Y_{\nu}} \sim \tilde{m} {\bf Y_{\nu}},
\end{eqnarray}
in which $F/M \sim \tilde{m}$ sets the scale of soft-breaking 
masses (where $\tilde{m} \sim {\rm TeV}$). There are also $D$ term contributions from the K\"{a}hler
potential, which are intrinsically nonanalytic.    These contributions lead to suppressed effective Yukawa couplings \begin{eqnarray}
\label{hardhol} 
\frac{1}{M^2} \left (\widehat{X}^\dagger \widehat{L} \cdot
\widehat{H}_u {\bf \bar{Y}_{\nu}}  \widehat{N} \right)_{D}\, = 
{L} \cdot {H}_u {\bf \widetilde{Y}_{\nu}} \nu_R^{c},
\end{eqnarray}
in which the effective Yukawa coupling is
\begin{eqnarray}
{\bf \widetilde{Y}_{\nu}} \equiv \frac{F}{M^2} {\bf \bar{Y}_{\nu}} \sim  \frac{\tilde{m}}{M} {\bf \bar{Y}_{\nu}},
\end{eqnarray}
which were previously studied \cite{neutrinosusybreaking1}.  They also lead to hard supersymmetry breaking effective fermion Yukawa couplings
of the wrong-Higgs form:
\begin{eqnarray}
\label{hard} \frac{1}{M^2}\left(\widehat{X}^\dagger \widehat{L} \cdot
\widehat{H}_d^{c} {\bf \bar{Y}^\prime_{\nu}}  \widehat{N}\right)_{D}\, = 
{L} \cdot {H}_d^{c} {\bf \widetilde{Y}^\prime_{\nu}}
\nu^c_R,
\end{eqnarray}
in which the effective Yukawa coupling is
\begin{eqnarray}
{\bf \widetilde{Y}^\prime_{\nu}} \equiv \frac{F}{M^2} {\bf \bar{Y}^\prime_{\nu}} \sim  \frac{\tilde{m}}{M} {\bf \bar{Y}^\prime_{\nu}}.
\end{eqnarray}
In addition to the usual scalar mass-squared terms,
\begin{eqnarray}
\label{mnu} \frac{1}{M^2} \left( \widehat{X}
\widehat{X}^{\dagger} \widehat{N}^{c} {\bf K}_{\nu} \widehat{N}
\right)_{D} \, = \, \widetilde{\nu}_R^{T} {\bf
m}_{\widetilde{N}}^2 \widetilde{\nu}_R^{c},
\end{eqnarray}
with
\begin{eqnarray}
 {\bf m}_{\widetilde{N}}^2 \equiv
\frac{F^2}{M^2} {\bf K}_{\nu} \sim \tilde{m}^2 {\bf K}_{\nu},
\end{eqnarray}
$D$ terms also lead to wrong-Higgs nonanalytic trilinear couplings:\footnote{Unlike the holomorphic couplings, the nonholomorphic couplings are independent of the superpotential.}
\begin{eqnarray}
\label{anup} \frac{1}{M^3} \left( \widehat{X}
\widehat{X}^{\dagger} \widehat{L}  \cdot \widehat{H}_d^{c} {\bf
Y}_{\nu}^{\prime} \widehat{N}\right)_{D} \, = \,
\widetilde{L}\cdot {H}_d^{c} {\bf A}^\prime_{\nu}
\widetilde{\nu}_R^{c},
\end{eqnarray}
with
\begin{eqnarray} {\bf
A}_{\nu}^{\prime} \equiv \frac{F^2}{M^3} {\bf
Y}_{\nu}^{\prime} \sim \frac{\tilde{m}^2}{M} {\bf Y}_{\nu}^{\prime}.
\end{eqnarray}
Hence, ${\bf A}_{\nu}^{\prime}$ is suppressed by $F/M^2=\tilde{m}/M$ with respect to ${\bf A}_{\nu}\sim F/M$.  It is the 
$F/M^2$ suppression which plays a key role in neutrino mass generation in both cases. The $F/M^2$ suppression has been discussed previously \cite{neutrinosusybreaking1,neutrinosusybreaking2}; however, these works present models in which nonholomorphic terms lead to Majorana masses and holomorphic operators lead to Dirac masses, and do not typically allow for the right-handed neutrinos to have nontrivial charges under additional gauge symmetries.
 
To allow the wrong-Higgs couplings of Eq.~(\ref{hard}) and Eq.~(\ref{anup}) and forbid the usual neutrino Yukawa couplings (both tree level and effective, as in Eq.~(\ref{anu}) and 
Eq.(\ref{hardhol})),  we assume that the right-handed neutrinos are charged under an extended gauge group. This prevents $\widehat{N}^i$ from acquiring a large tree-level Majorana mass\footnote{See \cite{Luhn:2007gq} for related work involving discrete gauge symmetries.} (in contrast to the seesaw mechanism), and has the added advantage that 
the nonanalytic trilinear couplings of Eq.~(\ref{anup}) now are soft breaking terms ({\it i.e.}, no quadratic divergences are induced in the scalar 
sector). The simplest gauging, though not the only logical possibility,  is to add a new Abelian gauge factor $U(1)^\prime$, with charges that satisfy
\begin{eqnarray}
\label{charge}
Q_L+ Q_{H_u} + Q_N &\neq& 0,\\
Q_L - Q_{H_d} + Q_N &=& 0.
\end{eqnarray}
These conditions are clearly inconsistent with having a bare superpotential $\mu$ term. The remedy is to replace the $\mu$ parameter by a chiral SM singlet $\widehat{S}$ with a nonvanishing $U(1)^\prime$ charge $Q_S$, with  $Q_S + Q_{H_u} + Q_{H_d}  = 0$ \cite{u1p}, such that an effective $\mu$ term is induced by the 
VEV of $S$.\footnote{One can also require that charged fermion masses are
generated radiatively, which requires much larger soft  trilinear couplings.}  
In this case, it is worth noting that upon $U(1)^\prime$ breaking, superpotential holomorphic couplings of the form  
\begin{equation}
\frac{1}{M}\widehat{S} \widehat{L} \cdot
\widehat{H}_u {\bf Y^{\prime\prime}_{\nu}}  \widehat{N}
\end{equation}
may also be generated. As discussed  in~\cite{higherd}, these
may give rise to an additional (``right-Higgs") contribution to the Dirac masses of  a similar order of magnitude:
%\footnote{See \cite{Luhn:2007gq} for  related work using discrete gauge symmetries.}
\begin{equation}
m_f=\frac{\langle S \rangle}{M}  {\bf Y^{\prime\prime}_{\nu}}\langle H^0_u \rangle \sim \frac{\tilde{m}}{M} {\bf Y^{\prime\prime}_{\nu}}\langle H^0_u \rangle.
\end{equation}
We assume any $U(1)^\prime$ gauge anomalies are cancelled by GUT remnants at the TeV scale; one can also consider anomaly free 
family-dependent $U(1)^{\prime}$ groups \cite{kane}.
\begin{figure}
\includegraphics[width=3in, height=1.5in]{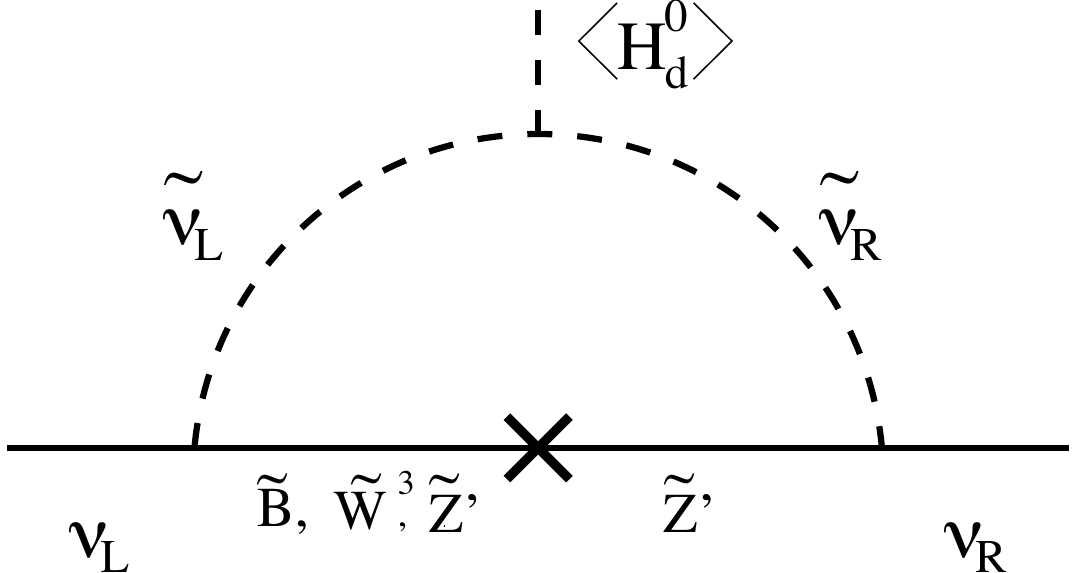}
\caption{\label{sekil1} The one-loop diagram that generates radiative Dirac neutrino masses. 
%The gaugino mass insertion,
%denoted by a cross, generates the requisite chirality violation. 
%The chiral flavor violation is provided by ${\bf A}_{\nu}^{\prime}$. 
%The right-handed neutrinos couple to $U(1)^{\prime}$
%gaugino $\widetilde{Z}^{\prime}$ only whereas the left-handed
%ones do also to neutral wino $\widetilde{W}^3$ and bino $\widetilde{B}$.
}
\end{figure}

We now turn to a more precise analysis of the neutrino masses generated by Eq.~(\ref{hard}) and Eq.~(\ref{anup}). The Yukawa interaction Eq.~(\ref{hard}) induces a Dirac neutrino mass 
\begin{eqnarray}
\label{treemass}
m_{\nu} = \langle H_d^0 \rangle {\widetilde{{\bf Y}}}_{\nu}^{\prime},
\end{eqnarray}
in agreement with  Eq.~(\ref{treemass1}) and Eq.~(\ref{estimate1}). This interaction is technically 
hard, but the resulting Higgs mass shift $\delta m_{H_d}^2 =   - (1/(8 \pi^2)) \widetilde{{\bf Y}}_{\nu}^{\prime \dagger} \widetilde{{\bf
Y}}_{\nu}^{\prime }  M^2  = - (1/(8 \pi^2)) \tilde{m}^2  \bar{{\bf 
Y}}_{\nu}^{\prime \dagger } \bar{{\bf
Y}}_{\nu}^{\prime }$
%$\delta m_{H_d}^2 \sim \left| {\widetilde{{\bf Y}}}_{\nu}^{\prime} \right|^2 M^2/(4 \pi)^2 \sim \tilde{m}^2 \left|{\bar{{\bf Y}}}_{\nu}^{\prime}\right|^2/(4 \pi)^2$,
is too small to leave any impact on the gauge hierarchy.

For the radiatively induced neutrino masses, the requisite Lagrangian terms are
\begin{eqnarray}
\label{etki}
\frac{g_Y}{\sqrt{2}}  \widetilde{\nu}_{L}^{\dagger}
{\cal N}_i \chi^{0}_{i} \nu_{L}
+ {\sqrt{2}} g_Y^{\prime} Q_N \widetilde{\nu}_R^{T}
N^0_{Z^{\prime} i} \chi^{0}_i \nu_R^c + \mbox{h.c.},
\end{eqnarray}
in which $N^0_{\eta^0 i}$ denotes the contamination of the neutralino gauge eigenstate $\eta^0 \in \{\widetilde{Z}^{\prime},
\widetilde{B}, \widetilde{W}^3, \widetilde{H}_d^0,
\widetilde{H}_u^0, \widetilde{S}\}$ in the $i$th neutralino $\chi^0_i$
($i=1,\dots,6$), and ${\cal N}_i$ is
 \begin{eqnarray}
{\cal N}_i=\cot \theta_{W} N^0_{W^3 i} - N^0_{B i} + 2 Q_L
\frac{g_Y^{\prime}}{g_{Y}} N^0_{Z' i}.
\end{eqnarray}
These interactions induce Dirac neutrino masses at one loop, as shown in Figure 1, of the form:\footnote{Due to small mixing, the $\widetilde{B}$ and $\widetilde{W}^3$ contributions are typically subdominant  to that of the $\widetilde{Z}^{\prime}$.}
\begin{eqnarray}
\label{mneut} {\bf m}_{\nu \, ab} &=&  \frac{g_Y
g_Y^{\prime} \langle H_d^0 \rangle Q_N}{32 \pi^2} 
%\sum_{c,d=1}^3 \sum_{i=1}^{6} 
\Bigg\{ {\cal{S}}_{L \, ac}
%\left 
({\cal{S}}_{L}^{\dagger} {\bf
A}_{\nu}^{\prime}{\cal{S}}_{R}%\right 
)_{cd}
{\cal{S}}_{R \,db}^{\dagger}\ \nonumber\\
&\times & m_{\chi_i^0} N^{0}_{Z^{\prime} i}{\cal N}_i
% \left(\cot \theta_{W}
%N^0_{W^3 i} - N^0_{B i} + 2 Q_L \frac{g_Y^{\prime}}{g_Y} N^0_{
%Z^{\prime}i}\right)\,
F\left (m_{\tilde{\nu}_{L\,c}}^2, m_{\tilde{\nu}_{R\,d}}^2,
m_{\chi^0_i}^2 \right )\Bigg\},
\end{eqnarray}
in which repeated indices are summed over, and ${\cal{S}}_L$ and
${\cal{S}}_R$ are the sneutrino mixing matrices,\footnote{Their 
mass-squares are obtained by adding the associated $D$-term contributions 
\begin{eqnarray}
{\bf m}_{\tilde{\nu}_L}^2 &=& {\bf m}_{\tilde{L}}^2 + \frac{1}{2} \cos 2\beta M_Z^2  + \frac{1}{2} Q_L \delta_{Z^{\prime}}^2\nonumber\\
{\bf m}_{\tilde{\nu}_R}^2 &=& {\bf m}_{\tilde{N}}^2 + \frac{1}{2} Q_N \delta_{Z^{\prime}}^2,
\end{eqnarray}
with $\delta_{Z^{\prime}}^2 = 2 {g_Y^{\prime}}^2 \left( Q_{H_u} \langle H_u^0 \rangle^2 + 
Q_{H_d} \langle H_d^0 \rangle^2 + Q_{S} \langle S \rangle^2 \right)$. $\langle S \rangle $ sets the effective $\mu$ parameter below the 
$U(1)^{\prime}$ breaking scale \cite{u1p}.} defined via
\begin{eqnarray}
{\cal{S}}_L^{\dagger} {\bf m}_{\tilde{\nu}_L}^2
{\cal{S}}_L&=&
\mbox{diag}.\left(m_{\tilde{\nu}_{L\, 1}}^2,
m_{\tilde{\nu}_{L\,2}}^2,
m_{\tilde{\nu}_{L\, 3}}^2\right)\\
{\cal{S}}_R^{T} {\bf m}_{\tilde{\nu}_R}^2
{\cal{S}}_R^{*} &=&
\mbox{diag}.\left(m_{\tilde{\nu}_{R\,1}}^2,
m_{\tilde{\nu}_{R\,2}}^2,
m_{\tilde{\nu}_{R\,3}}^2\right).
\end{eqnarray}
The loop function appearing in Eq.(\ref{mneut}) is given by
\begin{eqnarray}
F\left(m_1^2, m_2^2, m^2\right) = \frac{1}{m_1^2 - m_2^2} \left(
\frac{\ln \beta_1}{\beta_1 -1} - \frac{\ln \beta_2}{\beta_2
-1}\right).
\end{eqnarray}
$\beta_i = m^2/m_i^2$ reduces to $1/2 m^2$ when $m_1=m_2=m$.

Eq.~(\ref{mneut}) demonstrates that 
%$m_{\nu}$ is proportional to $g_Y^{\prime} Q_N$. In other words,  
neutrinos acquire
Dirac masses radiatively only if the right-handed
neutrinos are gauged under the $U(1)^\prime$ symmetry. 
$U(1)^\prime$ invariance thus not only ensures that the nonanalytic trilinear terms are 
soft, but also provides the the chirality flip required for neutrino mass generation through the $\widetilde{Z}^{\prime}$, which couples to both left- and right-handed neutrinos.
%Eq.~(\ref{etki}, %The size of the neutrino masses is controlled by a number of
%parameters associated with the sneutrino and neutralino sectors.
%Once these parameters are given, one can explicitly compute all
%the entries of $m_{\nu}$. 
%The nonanalytic trilinear coupling ${\bf
%A}_{\nu}^{\prime}$, with its built-in $\tilde{m}/M$
%suppression, plays an important role in suppressing the light neutrino masses as in Eq.~(\ref{estimate2}).
%Roughly, $m_{\nu} \sim 10^{-2} {\bf A}_{\nu}^{\prime}$ (assuming that all other parameters and ratios of the soft masses
%are ${\cal{O}}(1)$). This simple form gives $m_{\nu} \sim 10^{-3}\
%{\rm eV}$ in agreement with experimental limits when $M \sim
%M_{GUT}$. For gravity mediation, the messenger scale becomes $M
%\sim M_{Pl}$, and a prediction of the neutrino masses requires
%${\bf A}_{\nu}^{\prime}$ not to scale as $\tilde{m}^2/M$, but
%instead to have a larger value. 

For $M\sim M_{GUT}$, the neutrino masses are in the right range (the $\alpha/2\pi$ suppression can be countered by relaxing the degeneracy among 
the superpartner masses; this factor is absent for the tree-level masses of Eq.~(\ref{treemass})).  If $M\sim M_{Pl}$, 
an enhancement %in the nonanalytic terms
is required.  For other mediation mechanisms the messenger scale can be lowered, depending on the details of the model.
% In general,
%however, finding a supersymmetry breaking mechanism that naturally
%reproduces experimental data for not only the neutrino masses as
%given in Eq.~(\ref{mneut}), but also for other low-energy
%observables, depends largely on how the hidden sector is modelled,
%how flavor symmetry breaking is implemented, and how supersymmetry
%breaking is communicated to the observable sector.

%\footnote
 
The flavor structure of the tree-level Dirac neutrino mass Eq.~(\ref{treemass}) depends
only on ${\bf \bar{Y}}_{\nu}^{\prime}$ in Eq.(\ref{hard}). However, the flavor
structure of the radiative neutrino masses  involves  ${\bf m}_{\tilde{\nu}_L}^2$, ${\bf
m}_{\tilde{\nu}_R}^2$, and ${\bf A}_{\nu}^{\prime}$.
%(note that the scalar couplings ${\bf K}_{\nu}$, ${\bf Y}_{\nu}$, ${\bf Y}_{\nu}^{\prime}$, and 
%${\bf \bar{Y}}_{\nu}^{\prime}$ must be taken with care due to constraints from FCNC data).
% in Eq.~(\ref{anu}), Eq.~(\ref{mnu}), and 
%Eq.~(\ref{anup}) imply 
%are matrices in flavor space that % , and may signify nontrivial flavor interactions at the scale $M$.  Essentially, these matrices 
%parametrize our ignorance of the true mechanism that breaks flavor. These couplings must be taken with care, as stringent flavor-dependent limits result from low-energy FCNC data.)   
If the left-handed and right-handed sneutrinos are approximately degenerate in mass, the 
%Eq.~(\ref{mneut}) reduces to
%\begin{eqnarray}
%\left(m_{\nu}\right)_{a b} &=&  \frac{g_Y g_Y^{\prime}}{32 \pi^2}
%\langle H_d^0 \rangle Q_N \left({\bf A}_{\nu}^{\prime}\right)_{a b} \nonumber\\
%&\times & \sum_{i=1}^{6} m_{\chi_i^0} N^{0}_{Z^{\prime} i} {\cal N}_i
%F\left(m_{\tilde{\nu}_{L}}^2, m_{\tilde{\nu}_{R}}^2
%m_{\chi^0_i}^2\right),
%\end{eqnarray}
%in which case the 
neutrino mixings are controlled by the nonanalytic
trilinear coupling ${\bf A}_{\nu}^{\prime}$ alone. Alternatively, ${\bf A}_{\nu}^{\prime}$ may be strictly diagonal, such that neutrino mixings arise 
from nontrivial flavor structures of ${\bf m}_{\tilde{\nu}_L}^2$ and ${\bf m}_{\tilde{\nu_R}}^2$.
%The most important feature is that neutrino masses are linked to the
%the soft supersymmetry breaking sector of a low energy
%$U(1)^{\prime}$ model with right-handed neutrinos.

The radiative mechanism that leads to fermion masses also generically induces electric and
magnetic dipole moments \cite{shrock,soft1}. However, in this scenario, the neutrino dipole moments vanish at one loop.  This occurs because the right-handed neutrinos do not couple directly to the higgsinos through Yukawa interactions, and they do not have any charged gaugino with which to interact.   
Dirac neutrino masses also induce dipole moments within the SM of order $10^{-19} \mu_B$, which are much smaller than the best available bounds (of order $10^{-12} \mu_B$) \cite{dipole}.

In this Letter, we have discussed mechanisms to induce naturally suppressed neutrino Dirac masses within gauge-extended models with low energy supersymmetry.  Neutrino mass terms of the ``wrong-Higgs" type are generated either at tree level from formally hard (but in practice safe) effective Yukawa couplings, or radiatively due to nonanalytic soft supersymmetry breaking interactions.
% with a crucial role played by the nonanalytic soft terms  Eq.~(\ref{anup}).
The neutrino mass scale naturally falls within the experimentally
allowed range due to the $F/M^2 \sim \tilde{m}/M$ suppression.  
%In contrast to many scenarios,
%mechanisms for suppressing the overall neutrino mass scale, 
Moreover, this mechanism is operational for models in which the right-handed neutrinos are not complete singlets of the low energy gauge group.  This scenario, apart from providing an understanding of the
origin of naturally suppressed Dirac neutrino masses, allows for a natural resolution of the supersymmetric $\mu$ problem and leads to TeV-scale $U(1)^\prime$ physics which
should be testable at forthcoming colliders such as the LHC.

\acknowledgments
D.D. is supported by  Alexander von Humboldt-Stiftung
Friedrich Wilhelm Bessel-Forschungspreise and by the Turkish Academy of 
Sciences via the GEBIP grant. L.E.  is supported by the U.S. Department of Energy grant DE-FG02-95ER40896.  P.L. is supported by Friends of the IAS and by the NSF grant PHY-0503584.  We thank D.~J.~Chung, A.~Ibarra, I.~W.~Kim, C.~Luhn, N.~Seiberg, and L.~T.~Wang for fruitful discussions.

%%%%%%%%%%%%%%%%%%%%%%%%%%%%%%%%%%%%%%%%%%%%%%%%%%%%%%%%%%%%%%%%%%%%%%%%%%%

\end{document}